\documentclass[letterpaper, 10 pt, conference]{ieeeconf}
\IEEEoverridecommandlockouts
\usepackage[utf8]{inputenc}
\usepackage{amsmath}
\usepackage{mathrsfs}
\usepackage{amssymb}
\usepackage{color}
\usepackage{dsfont}
\usepackage{cite}
\usepackage{comment}
\usepackage{float}
\usepackage{graphicx}
\usepackage[font = footnotesize]{caption}
\usepackage[font = footnotesize]{subcaption}
\usepackage{tikz}
\usepackage{booktabs}
\usepackage{tabularx}
\usepackage{multirow}
\usepackage{bm}
\usepackage{makecell}
\usepackage{indentfirst}
\newcolumntype{C}[1]{>{\centering\let\newline\\\arraybackslash\hspace{0pt}}m{#1}}

\usepackage{color, colortbl}
\definecolor{Yellow}{rgb}{1,1,0}

\usepackage{amsmath} 
\usepackage{amssymb}  
\usepackage{amsfonts}

\newtheorem{proposition}{Proposition}[section]

\usepackage{graphicx}
\usepackage{epstopdf}
\usepackage{float} 
\usepackage[skip=0pt,font=scriptsize]{caption}

\usepackage[margin = 1in]{geometry}

\usepackage{etaremune}

\usepackage{tikz}
\usetikzlibrary{calc,patterns,decorations.pathmorphing,decorations.markings}
\usetikzlibrary{shapes,arrows}
\usepackage{verbatim}

\tikzstyle{block} = [draw, fill=blue!20, rectangle, 
    minimum height=3em, minimum width=6em]
\tikzstyle{sum} = [draw, fill=blue!20, circle, node distance=1cm]
\tikzstyle{input} = [coordinate]
\tikzstyle{output} = [coordinate]
\tikzstyle{pinstyle} = [pin edge={to-,thin,black}]

\usetikzlibrary{positioning}
\usetikzlibrary{math}

\usepackage{tikz}
\usetikzlibrary{shapes,arrows,calc,positioning}
\tikzstyle{bigblock} = [draw, fill=blue!20, rectangle, 
    minimum height=6em, minimum width=8em]
\tikzstyle{medblock} = [draw, fill=blue!20, rectangle, 
    minimum height=4em, minimum width=4em]    
\tikzstyle{mux} = [draw, fill=black!20, rectangle, 
    minimum height=5em, minimum width=0.1em]    
\tikzstyle{smallblock} = [draw, fill=blue!20, rectangle, 
    minimum height=3em, minimum width=4em]
\tikzstyle{sum} = [draw, fill=blue!20, circle, node distance=1cm]
\tikzstyle{signal} = [coordinate]
\tikzstyle{pinstyle} = [pin edge={to-,thin,black}]
\tikzstyle{block} = [draw, fill=blue!20, rectangle, 
    minimum height=3em, minimum width=6em]
\tikzstyle{blockS} = [draw, fill=blue!20, rectangle, 
    minimum height=3em, minimum width=4em]    
\tikzstyle{input} = [coordinate]
\tikzstyle{output} = [coordinate]
\usetikzlibrary{matrix}

\tikzset{add/.style n args={4}{
    minimum width=1mm,
    path picture={
        \draw[black, thick] 
            (path picture bounding box.south east) -- (path picture bounding box.north west)
            (path picture bounding box.south west) -- (path picture bounding box.north east);
        }
    }
}

\tikzset{Frame_into/.pic={
        code={\tikzset{scale=1}
        \tikzmath
            {
                \l  = 1;
                \Rc = 0.15;
            } 
        \draw [thick,->] (0, 0) -- +(\l, 0);
        \draw [thick,->] (0, 0) -- +(0, \l);
        \draw [thick, fill=white] 
			    (0,0) circle [radius=\Rc];
	    \draw [thick] ({\Rc*cos(45)}, {0\Rc*sin(45)}) 
	               -- ({\Rc*cos(225)}, {0\Rc*sin(225)});
        \draw [thick] ({\Rc*cos(135)}, {0\Rc*sin(135)}) 
	               -- ({\Rc*cos(315)}, {0\Rc*sin(315)});
  }}
}

\tikzset{Frame_outof/.pic={
        code={\tikzset{scale=1}
        \tikzmath
            {
                \l  = 1;
                \Rc = 0.15;
            } 
        \draw [thick,->] (0, 0) -- +(\l, 0);
        \draw [thick,->] (0, 0) -- +(0, \l);
        \draw [thick, fill=white] 
			    (0,0) circle [radius=\Rc];
	    \draw [thick, fill=black] 
			    (0,0) circle [radius={0.2*\Rc}];
  }}
}

\usepackage{hyperref}
\usepackage{xcolor}
\hypersetup{
    colorlinks,
    linkcolor={blue!100!black},
    citecolor={blue!50!black},
    urlcolor={blue!80!black}
}

\newcommand{\bc}{\begin{center}}
\newcommand{\ec}{\end{center}}
\newcommand{\benum}{\begin{enumerate}}
\newcommand{\eenum}{\end{enumerate}}
\newcommand{\nn}{\nonumber}
\newcommand{\matl}{\left[ \begin{array}}
\newcommand{\matr}{\end{array} \right]}
\newcommand{\matls}{\left[ \begin{smallmatrix}}
\newcommand{\matrs}{\end{smallmatrix} \right]}
\newcommand{\isdef}{\stackrel{\triangle}{=}}

\newcommand{\inv}{^{-1}}

\newcommand{\vect}[1]{\overset{\rightharpoonup}{#1}}

\newcommand{\rmA}{{\rm A}}

\newcommand{\rmT}{{\rm T}}

\newcommand{\rmc}{{\rm c}}
\newcommand{\rmd}{{\rm d}}

\newcommand{\rmf}{{\rm f}}

\newcommand{\rmi}{{\rm i}}

\newcommand{\rmp}{{\rm p}}

\newcommand{\BBR}{{\mathbb R}}

\renewcommand{\matl}{\begin{bmatrix}}
\renewcommand{\matr}{\end{bmatrix} }

\usepackage{hhline}

\usetikzlibrary{shapes,arrows,calc,positioning}
\tikzstyle{bigblock} = [draw, fill=blue!20, rectangle, 
    minimum height=6em, minimum width=8em]
\tikzstyle{medblock} = [draw, fill=blue!20, rectangle, 
    minimum height=4em, minimum width=4em]    
\tikzstyle{mux} = [draw, fill=black!20, rectangle, 
    minimum height=5em, minimum width=0.1em]    
\tikzstyle{smallblock} = [draw, fill=blue!20, rectangle, 
    minimum height=3em, minimum width=4em]
\tikzstyle{sum} = [draw, fill=blue!20, circle, node distance=1cm]
\tikzstyle{signal} = [coordinate]
\tikzstyle{pinstyle} = [pin edge={to-,thin,black}]
\tikzstyle{block} = [draw, fill=blue!20, rectangle, 
    minimum height=3em, minimum width=6em]
\tikzstyle{blockS} = [draw, fill=blue!20, rectangle, 
    minimum height=3em, minimum width=4em]  
\tikzstyle{sum} = [draw, fill=blue!20, circle, node distance=1.5cm]
\tikzstyle{gain} = [draw, fill=blue!20, regular polygon, regular polygon sides = 3, node distance=1.25cm, shape border rotate = -90]
\tikzstyle{mult} = [draw, fill=blue!20, circle, node distance=1.25cm ,inner sep=0pt, minimum size = 0.3cm]

\tikzstyle{input} = [coordinate]
\tikzstyle{output} = [coordinate]

\newcounter{example}

\title{\LARGE Continuous-Time Retrospective Cost Adaptive Control \\ for Stabilization and Tracking with Experimental Results}

\title{\LARGE Continuous-Time Output Feedback Adaptive Control \\ for Stabilization and Tracking with Experimental Results}

\author{\large Mohammad Mirtaba and Ankit Goel
\thanks{Mohammad Mirtaba and Ankit Goel are with the Department of Mechanical Engineering, University of Maryland, Baltimore County, MD 21250.
{\tt \small \{mmirtab1,ankgoel\}@umbc.edu}}
}

\begin{document}

\maketitle

\begin{abstract}
This paper presents a continuous-time output feedback adaptive control technique for stabilization and tracking control problems. 
The adaptive controller is motivated by the classical discrete-time retrospective cost adaptive control algorithm.
The particle swarm optimization framework automates the adaptive algorithm's hyperparameter tuning. 
The proposed controller is numerically validated in the tracking problems of a double integrator and a bicopter system and is experimentally validated in an attitude stabilization problem.
Numerical and experimental results show that the proposed controller is an effective technique for model-free output feedback control.

\end{abstract}

\section{Introduction}
\label{sec:introduction}





Adaptive control has widespread applications across various industries, including aerospace \cite{litre1}, automotive, robotics \cite{litre2}, and process control. 
Unlike model-based control design techniques, which rely on prior knowledge of the model of the system to design a controller, adaptive controllers dynamically adjust their controller parameters in real-time to achieve and maintain desired performance.
Due to their ability to adjust dynamically, adaptive control is also suitable for systems with uncertain or time-varying parameters. 
Furthermore, this capability makes adaptive control particularly valuable in applications where system dynamics are not fully known or are subject to changes due to environmental variations, component wear, etc.

A foundational approach in adaptive control is the Model Reference Adaptive Control (MRAC) \cite{bookmrac}, where the error between a reference model's state and the actual system state is used to drive the adaptation laws to yield stabilizing controller.
$L_1$ adaptive control, introduced in \cite{L1}, uses a similar approach to design the adaptive laws to construct stabilizing controllers. 
However, in both of these approaches, the full state measurement is required to drive the adaptation law. 
Furthermore, both approaches require an apriori stabilizing controller \cite{loneapp,loneapp2}. 
%
%
%
Gain scheduling and auto-tuning methods are another popular paradigm in adaptive control. \cite{fuzzy1,fuzzy2} use fuzzy logic for PID gain scheduling where membership functions based on error and derivative of error are defined and gains are calculated with fuzzy logic.
With recent advances in machine learning, reinforcement learning-based PID gain scheduling is also investigated \cite{rlpid1, rlpid2}.
However, these techniques often require large computational resources and multiple simulation iterations.
Furthermore, such techniques also suffer from the sim-to-real gap if the controller is synthesized in an offline framework. 


Retrospective cost adaptive control (RCAC) is a discrete-time adaptive output feedback control technique for stabilization, command following, and disturbance rejection problems.
RCAC optimizes a linearly parameterized controller using retrospective cost optimization \cite{rcac1,  rcac3}.
Over the last two decades, the discrete-time RCAC algorithm has been applied to several control problems, including multicopter control \cite{rcacapp1}, flight control \cite{rcacapp2, AnsariJAIA}, combustion control \cite{goel2018}, noise control \cite{xie2017spatial}, etc. 
However, since the classical RCAC algorithm is designed for discrete-time systems, its application to continuous-time systems requires either the discrete sampling or the discretization of the continuous-time system, thus making sampling time a key performance-affecting factor.
Furthermore, the stability analysis of the closed loop becomes challenging due to the mixing of a discrete-time controller and a continuous-time plant. 
%

The paper is focused on the continuous-time extension of the classical RCAC algorithm, denoted henceforth by CTRCAC.
The contributions of the paper are 1) the extension of the multi-input, multi-output extension of the CTRCAC as presented in \cite{dai2016continuous}, 2) the introduction of time-domain filtering using state-space form of the filter, and 3) the construction of various controller parameterization. 
CTRCAC is then applied to the command-following problem in a double integrator system, the trajectory-tracking problem in a bicopter system, and finally to the attitude control of a rigid body. 
Numerical simulations and physical experiments show that CTRCAC, like classical RCAC, is an effective technique for model-free output feedback adaptive control of dynamic systems. 
However, the continuous-time formulation can be directly integrated with either continuous-time models or physical systems in the loop without discretizing the system or the controller dynamics.
Furthermore, in this paper, we use a particle-swarm optimizer to automate the tuning of CTRCAC hyperparameters, thus reducing the need to tune them manually.




The paper is organized as follows.
Section \ref{sec:CTRCAC} reviews the continuous-time retrospective cost adaptive control presented in \cite{dai2016continuous} and extends it to the multi-input, multi-output case.
Section \ref{sec:applications} presents three applications of CTRCAC to the stabilization and tracking problems.
Finally, Section \ref{sec:conclusions} summarizes the work presented in this paper. 

\section{Continuous Time Retrospective Cost Adaptive Control}
\label{sec:CTRCAC}
This section briefly describes the continuous-time retrospective cost adaptive control and derives the update equations for the adaptive control law.
The algorithm presented in this paper extends the algorithm presented in \cite{dai2016continuous} by 
generalizing the algorithm to multi-input, multi-output systems and   
introducing state-space filtering.


Consider a dynamic system 
\begin{align}
    \dot x(t) &= f(x(t),u(t)), \\
    y(t) &= g(x(t)),
\end{align}
where 
$x(t) \in \BBR^{l_x}$ is the state, 
$u(t) \in \BBR^{l_u}$ is the input, 
$y(t) \in \BBR^{l_y}$ is the measured output, and 
the vector functions $f$ and $g$ are the dynamics and the output maps. 
Define the \textit{performance variable}
\begin{align}
    z(t) \isdef y(t) - r(t), 
    \label{eq:z_def}
\end{align}
where $r(t)$ is the reference signal. 
The objective is to design an adaptive output feedback control law to ensure that $z(t) \to 0.$

Consider a linearly parameterized control law 
\begin{align}
    u(t) 
        =
            \Phi(t) \theta(t), 
    \label{eq:u_para}
\end{align}
where the regressor matrix $\Phi(t) \in \BBR^{l_u \times l_\theta}$ contains the measured data and 
the vector $\theta(t) \in \BBR^{l_\theta} $  contains the controller gains to be optimized. 
Various linear parameterizations of MIMO controllers are described in \cite{goel_2020_sparse_para}.
Linear parameterizations of an $n$-th order transfer function and a PID controller are briefly reviewed in Appendix \ref{appndx:ControllerPara}. 

Next, using \eqref{eq:z_def}, define the \textit{retrospective performance}
\begin{align}
    \hat z(t, \hat \theta) 
        &\isdef
            z(t) + \Phi_\rmf(t) \hat \theta -u_\rmf(t),
\end{align}
where $\hat \theta$ is the controller gain to be optimized and  
the filtered regressor $\Phi_\rmf(t)$ and the filtered control $u_\rmf(t)$ are defined as
\begin{align}
    \Phi_\rmf(t) &\isdef G_\rmf(s) \left[ \Phi(t) \right] \in \BBR^{l_y \times l_\theta},
    \\
    u_\rmf(t) &\isdef G_\rmf(s) \left[ u(t) \right] \in \BBR^{l_y \times l_\theta},
\end{align}
where $G_\rmf(s)$ is a dynamic filter transfer function.


The filtered signals $\Phi_\rmf(t)$ and $u_\rmf(t)$ are computed in the time domain, as shown below. 
Let $(A_\rmf,B_\rmf,C_\rmf,D_\rmf)$ be a realization of $G_\rmf(s).$
Then, 
\begin{align}
    \dot x_\Phi &= A_\rmf x_\phi + B_\rmf \Phi, \\
    \Phi_\rmf &= C_\rmf x_\phi + D_\rmf \Phi,
\end{align}
and
\begin{align}
    \dot x_u &= A_\rmf x_u + B_\rmf u, \\
    u_\rmf &= C_\rmf x_u + D_\rmf u.
\end{align}

Next, define the retrospective cost 
\begin{align}
    J(t, \hat \theta)
        &=
            \int_{0}^{t} 
            \Bigg(
                \hat z(\tau, \hat \theta)^\rmT R_z \hat z(\tau, \hat \theta) 
                \nn \\ &\quad +
                (\Phi(\tau)   \hat \theta)^\rmT R_{u} (\Phi(\tau)  \hat \theta)
            \Bigg)
            \rmd \tau
            + \hat \theta^T R_\theta \hat \theta,
    \label{eq:J_RCAC}
\end{align}
where $R_z,
R_u,$ and $R_\theta$ are positive definite weighting matrices of appropriate dimensions.

\begin{proposition}
    Consider the cost function $J(t, \hat \theta)$ given by \eqref{eq:J_RCAC}.
    For all $t \ge 0, $ define the minimizer of $J(t, \hat \theta)$ by
    \begin{align}
        \theta(t)
            \isdef 
                \underset{\hat \theta \in \BBR^{l_\theta}}{\operatorname{argmin}} 
                J(t, \hat \theta).
    \end{align}
    Then, for all $t \ge 0, $ the minimizer satisfies
    \begin{align}
        \dot \theta
        &=
            -P \Phi_\rmf^\rmT R_z (z+ \Phi_\rmf  \theta -u_\rmf  )
            -P \Phi^\rmT R_{u} \Phi \theta, 
        \label{eq:theta_eq}
        \\
        \dot P
        &=
            -P
            \left(
            \Phi_\rmf ^\rmT 
                    R_z
                    \Phi_\rmf  
                    +
                    \Phi^\rmT R_{u} \Phi
            \right)
            P
    \end{align}
    where $P(0) = R_\theta^{-1}$ and $\theta(0) = 0.$
    
\end{proposition}

\begin{proof}
    Note that the cost function \eqref{eq:J_RCAC} can be written as
\begin{align}
    J(t, \hat \theta) 
        &=
            \hat \theta^T A(t) \hat \theta + 2 \hat \theta^T b(t) + c(t),
    \label{eq:J_reformulated}
\end{align}
where
\begin{align}
    A(t) 
        &\isdef
            \int_0^{t} 
            \left(
                    \Phi_\rmf ^\rmT 
                    R_z
                    \Phi_\rmf  
                    +
                    \Phi^\rmT R_{u} \Phi 
            \right)
            \rmd \tau + R_\theta,
    \nn
    \\
    b(t) 
        &\isdef
            \int_0^{t}
            \Phi_\rmf^\rmT
                R_z 
                (z-u_\rmf) \rmd \tau, 
    \nn \\
    c(t) 
        &\isdef
            \int_0^{t}
            (z-u_\rmf)^\rmT
                R_z 
                (z-u_\rmf)  
                \rmd \tau.
    \nn
\end{align}
Note that, for all $t\ge 0,$ $A(t)$ is positive definite, by construction. 
It follows from the Leibniz integral formula \cite{protter1985differentiation} that the matrices $A(t)$ and $b(t)$ satisfy
\begin{align}
    \dot A
        &=
            \Phi_\rmf ^\rmT 
                    R_z
                    \Phi_\rmf  
                    +
                    \Phi^\rmT R_{u} \Phi,  \nn 
    \\
    \dot b
        &=
            \Phi_\rmf^\rmT
                R_z 
                (z-u_\rmf) , \nn 
\end{align}
where $A(0) = R_\theta$ and $b(0) = 0.$


Next, define, for all $\ge 0,$ $P(t) \isdef A(t)^{-1}.$
Using the fact that $P(t) A(t) = I,$ it follows that
\begin{align}
    \dot P(t) = - P(t) \dot A(t) P(t), \nn
\end{align}
and thus 
\begin{align}
    \dot P 
        =
            -P
            \left(
            \Phi_\rmf ^\rmT 
                    R_z
                    \Phi_\rmf  
                    +
                    \Phi^\rmT R_{u} \Phi
            \right)
            P. \nn
\end{align}

Finally, note that the minimizer of \eqref{eq:J_reformulated} is given by
\begin{align}
    \theta(t)
        =
            A(t)^{-1} b(t)
        =
            P(t) b(t), \nn
\end{align}
and thus 
\begin{align}
    \dot \theta(t) 
        &=
            -\dot P b - P \dot b
        \nn 
        \\
        &=
            -P
            \left(
            \Phi_\rmf ^\rmT 
                    R_z
                    \Phi_\rmf  
                    +
                    \Phi^\rmT R_{u} \Phi
            \right) P
            P\inv \theta
            \nn \\ &\quad -
            P \Phi_\rmf^\rmT R_z (z-u_\rmf)
        \nn \\
        &=
            -P \Phi_\rmf^\rmT R_z (z+ \Phi_\rmf  \theta -u_\rmf  )
            -P \Phi^\rmT R_{u} \Phi \theta.
        \nn
\end{align}
\end{proof}

The control law is thus
\begin{align}
    u(t) = \Phi(t) \theta(t), 
\end{align}
where $\theta(t)$ is given by \eqref{eq:theta_eq}.
\section{Stabilization and Tracking Applications}
\label{sec:applications}

In this section, we apply the CTRCAC developed in the previous section to several stabilization and tracking problems. 
In particular, we consider a double integrator, a bicopter system, and a 3-dimensional attitude control problem. 
The double integrator and the bicopter system are simulated in Matlab and the application of CTRCAC is investigated in a simulation framework. 
The 3-dimensional attitude control problem considered in this paper consists of a Holybro X500 quadcopter frame mounted on a 3-2-1 gimbal platform that allows independent rotation of the three Euler angles, where the problem is to stabilize the attitude of the quadcopter.
The CTRCAC algorithm is implemented using the \textit{Matlab UAV toolbox package for PX4 autopilot} on a Pixhawk flight controller. 
Table \ref{tab:exmp_list} lists the examples considered in this paper. 

\begin{table}[h]
    \centering
    \renewcommand{\arraystretch}{1.5}
    \begin{tabular}{|p{0.075\textwidth}|p{0.075\textwidth}|p{0.075\textwidth}|p{0.075\textwidth}|}
        \hline
        Controller & Double $\quad$ Integrator & Bicopter & X500 
        \\ \hline
        Transfer function & Figure \ref{fig:crcac_2nd_doublei_data} & - & - 
        \\ \hline
        PID & Figure \ref{fig:crcac_PID_doublei_data} & Figure \ref{fig:bicopter_trajectory_data_pid} & -
        \\ \hline
        Cascaded PPI & Figure \ref{fig:crcac_ppi_doublei_data} & Figure \ref{fig:bicopter_trajectory_data_ppi} & Figure \ref{fig:platform_exp}
        \\ \hline
        FSFI & Figure \ref{fig:crcac_fsfi_doublei_data} & - & Figure \ref{fig:platform_exp}
        \\ \hline
    \end{tabular}
    \caption{List of examples in this paper.}
    \label{tab:exmp_list}
\end{table}

\subsection{Double Integrator}
Consider the double integrator  
\begin{align}
    \ddot q &= u, \\
    y &= q, 
\end{align}
where $q \in \BBR.$
The objective is to design an output-feedback adaptive control law to follow a unit step command $r(t) = 1.$
%

A particle swarm optimization (PSO) framework is used to optimize the RCAC hyperparameters $P_0$ and $G_\rmf(s).$ 
In particular, the filter $G_\rmf(s)$ is a first-order transfer function
\begin{align}
    G_\rmf(s)
        =
            \frac{1}{s+p_\rmf},
\end{align}
where PSO optimizes $p_\rmf$. 
In PSO, the lower and upper bound for $P_0$ is $10^{-4}$ and $10^4,$ respectively and 
and for $p_\rmf$ is $0.1$ to $10,$ respectively. 
The swarm size is set to $5.$

\begin{figure}[h]
    \centering
    \includegraphics[width=\columnwidth]{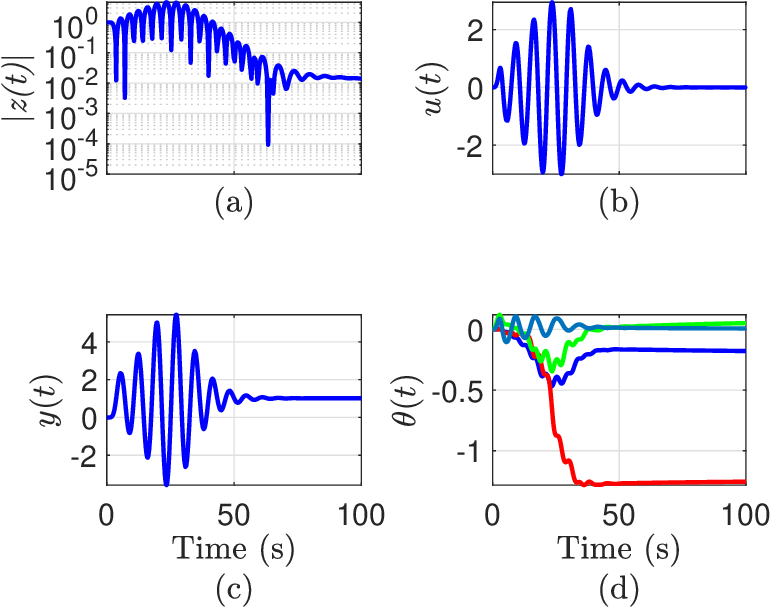}
    \caption{\textbf{[Double integrator.]} Closed-loop step response with a second-order controller optimized by RCAC.
    (a) shows the absolute value of the performance variable $z(t)$ on a logarithmic scale, 
    (b) shows the control $u(t)$ computed by the adaptive controller, 
    (c) shows the output $y(t)$, and
    (d) shows the controller gains $\theta(t)$ optimized by RCAC.}
    \label{fig:crcac_2nd_doublei_data}
\end{figure}

First, we optimize a second-order controller in a basic servo loop architecture to track the command. 
PSO is used to optimize the RCAC hyperparameters $P_0$ and $G_\rmf(s)$ as described above. 
The PSO optimized RCAC hyperparameters are $P_0 = 10^{0.6}$ and $p_\rmf = 8.15.$
Figure \ref{fig:crcac_2nd_doublei_data} shows the closed-loop response of the double integrator with the second-order controller optimized by RCAC. 

\begin{figure}[h]
    \centering
    \includegraphics[width=\columnwidth]{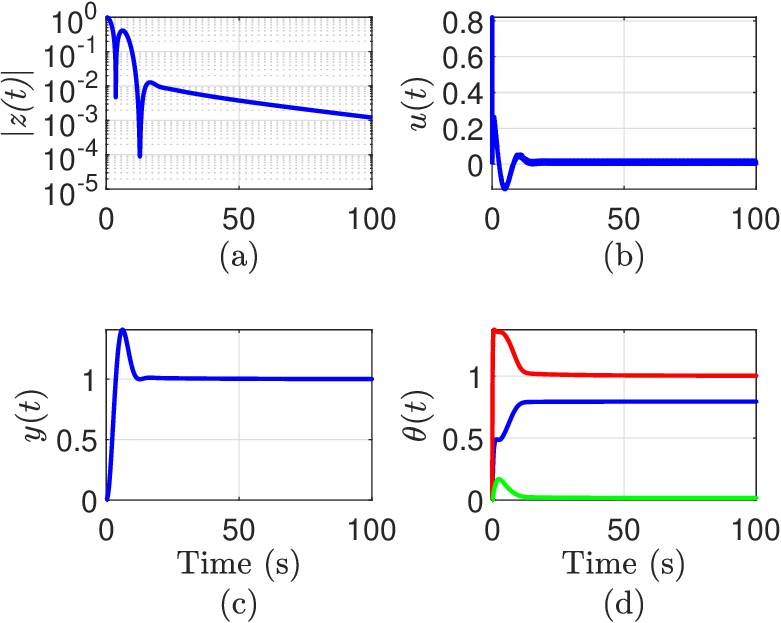}
    \caption{\textbf{[Double integrator.]} Closed-loop step response with a PID controller optimized by RCAC.
    (a) shows the absolute value of the performance variable $z(t)$ on a logarithmic scale, 
    (b) shows the control $u(t)$ computed by the adaptive controller, 
    (c) shows the output $y(t)$, and
    (d) shows the controller gains $\theta(t)$ optimized by RCAC.}
    \label{fig:crcac_PID_doublei_data}
\end{figure}

Next, we optimize a PID controller in a basic servo loop architecture to track the command.  
PSO is used to optimize the RCAC hyperparameters $P_0$ and $G_\rmf(s)$ as described above. 
The PSO optimized RCAC hyperparameters are $P_0 = 10^{-1.02}$ and $p_\rmf = 0.6508.$
Figure \ref{fig:crcac_PID_doublei_data} shows the closed-loop response of the double integrator with the second-order controller optimized by RCAC. 

\begin{figure}[h]
    \centering
    \includegraphics[width=\columnwidth]{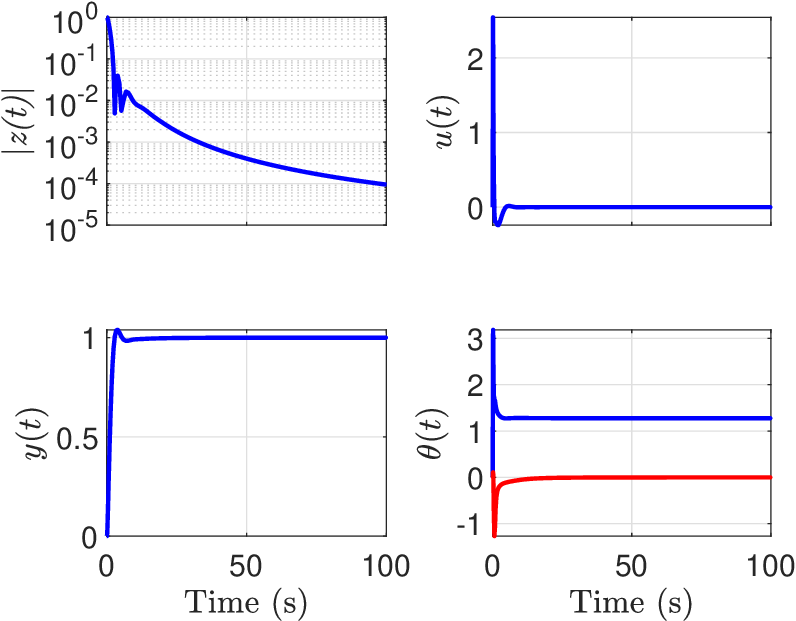}
    \caption{\textbf{[Double integrator.]} Closed-loop step response with a PPI controller optimized by RCAC.
    (a) shows the absolute value of the performance variable $z(t)$ on a logarithmic scale, 
    (b) shows the control $u(t)$ computed by the adaptive controller, 
    (c) shows the output $y(t)$, and
    (d) shows the controller gains $\theta(t)$ optimized by RCAC.}
    \label{fig:crcac_ppi_doublei_data}
\end{figure}

Next, we optimize the controllers in the cascaded architecture to track the command in the architecure described in Appendix \ref{sec:tracking}.
In this cascaded architecture, the outer loop consists of a proportional (P) controller, which allows the specification of a desired time constant and, thus, the desired transient response, and the inner loop consists of a proportional-integral (PI) controller to stabilize the plant. 
In this example, we set the proportional gain to unity, which yields a time constant of 1 second.
The PI controller in the inner loop is then synthesized by CTRCAC.
PSO is used to optimize the RCAC hyperparameters $P_0$ and $G_\rmf(s)$ as described above. 
The PSO optimized RCAC hyperparameters are $P_0 = 10^{-3.376}$ and $p_\rmf = 4.455.$
Figure \ref{fig:crcac_ppi_doublei_data} shows the closed-loop response of the double integrator with the FSFI controller optimized by RCAC. 

\begin{figure}[h]
    \centering
    \includegraphics[width=\columnwidth]{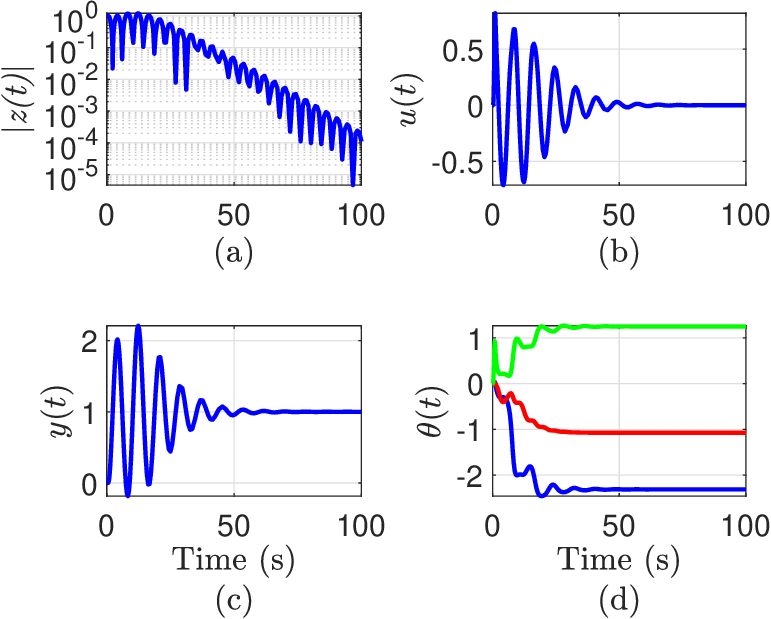}
    \caption{\textbf{[Double integrator.]} Closed-loop step response with a FSFI controller optimized by RCAC.
    (a) shows the absolute value of the performance variable $z(t)$ on a logarithmic scale, 
    (b) shows the control $u(t)$ computed by the adaptive controller, 
    (c) shows the output $y(t)$, and
    (d) shows the controller gains $\theta(t)$ optimized by RCAC.}
    \label{fig:crcac_fsfi_doublei_data}
\end{figure}

Finally, we optimize a full state feedback controller with integral action in the architecure described in Appendix \ref{sec:tracking}. 
PSO is used to optimize the RCAC hyperparameters $P_0$ and $G_\rmf(s)$ as described above. 
The PSO optimized RCAC hyperparameters are $P_0 = 10^{-1.278}$ and $p_\rmf = 3.314.$
Figure \ref{fig:crcac_fsfi_doublei_data} shows the closed-loop response of the double integrator with the FSFI controller optimized by RCAC. 

\subsection{Bicopter}
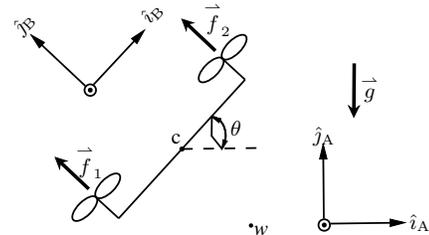
\begin{figure}[!ht]
    \centering
    \resizebox{0.75\columnwidth}{!}{
    
\tikzset{every picture/.style={line width=0.75pt}} 

\begin{tikzpicture}[x=0.75pt,y=0.75pt,yscale=-0.7,xscale=0.7]

\draw   (382.73,198.06) .. controls (384.44,196.15) and (387.38,195.98) .. (389.3,197.69) .. controls (391.22,199.41) and (391.38,202.35) .. (389.67,204.26) .. controls (387.96,206.18) and (385.02,206.35) .. (383.1,204.63) .. controls (381.18,202.92) and (381.02,199.98) .. (382.73,198.06) -- cycle ;
\draw    (385.37,137.84) -- (385.88,195.97) ;
\draw [shift={(385.35,135.84)}, rotate = 89.5] [fill={rgb, 255:red, 0; green, 0; blue, 0 }  ][line width=0.08]  [draw opacity=0] (12,-3) -- (0,0) -- (12,3) -- cycle    ;
\draw    (449.36,199.25) -- (391.23,200.02) ;
\draw [shift={(451.36,199.23)}, rotate = 179.24] [fill={rgb, 255:red, 0; green, 0; blue, 0 }  ][line width=0.08]  [draw opacity=0] (12,-3) -- (0,0) -- (12,3) -- cycle    ;
\draw  [fill={rgb, 255:red, 0; green, 0; blue, 0 }  ,fill opacity=1 ] (384.93,201.16) .. controls (384.93,200.46) and (385.5,199.9) .. (386.2,199.9) .. controls (386.9,199.9) and (387.47,200.46) .. (387.47,201.16) .. controls (387.47,201.86) and (386.9,202.43) .. (386.2,202.43) .. controls (385.5,202.43) and (384.93,201.86) .. (384.93,201.16) -- cycle ;
\draw    (322.35,83.75) -- (220.35,195.75) ;
\draw [shift={(271.35,139.75)}, rotate = 132.32] [color={rgb, 255:red, 0; green, 0; blue, 0 }  ][fill={rgb, 255:red, 0; green, 0; blue, 0 }  ][line width=0.75]      (0, 0) circle [x radius= 1.34, y radius= 1.34]   ;
\draw    (200.75,177.75) -- (220.35,195.75) ;
\draw   (183.73,197.05) .. controls (181.5,194.8) and (183.84,188.88) .. (188.95,183.81) .. controls (194.05,178.75) and (200,176.46) .. (202.22,178.71) .. controls (204.45,180.95) and (202.11,186.88) .. (197.01,191.94) .. controls (191.9,197.01) and (185.95,199.29) .. (183.73,197.05) -- cycle ;
\draw   (202.22,178.71) .. controls (200,176.46) and (202.33,170.54) .. (207.44,165.47) .. controls (212.55,160.41) and (218.49,158.12) .. (220.72,160.37) .. controls (222.94,162.61) and (220.61,168.54) .. (215.5,173.6) .. controls (210.39,178.66) and (204.45,180.95) .. (202.22,178.71) -- cycle ;
\draw    (302.75,65.75) -- (322.35,83.75) ;
\draw   (285.73,85.05) .. controls (283.5,82.8) and (285.84,76.88) .. (290.95,71.81) .. controls (296.05,66.75) and (302,64.46) .. (304.22,66.71) .. controls (306.45,68.95) and (304.11,74.88) .. (299.01,79.94) .. controls (293.9,85.01) and (287.95,87.29) .. (285.73,85.05) -- cycle ;
\draw   (304.22,66.71) .. controls (302,64.46) and (304.33,58.54) .. (309.44,53.47) .. controls (314.55,48.41) and (320.49,46.12) .. (322.72,48.37) .. controls (324.94,50.61) and (322.61,56.54) .. (317.5,61.6) .. controls (312.39,66.66) and (306.45,68.95) .. (304.22,66.71) -- cycle ;
\draw  [dash pattern={on 4.5pt off 4.5pt}]  (331.49,138.96) -- (271.35,139.75) ;
\draw  [fill={rgb, 255:red, 0; green, 0; blue, 0 }  ,fill opacity=1 ] (326.07,201.55) .. controls (326.07,200.98) and (326.53,200.51) .. (327.1,200.51) .. controls (327.68,200.51) and (328.14,200.98) .. (328.14,201.55) .. controls (328.14,202.12) and (327.68,202.59) .. (327.1,202.59) .. controls (326.53,202.59) and (326.07,202.12) .. (326.07,201.55) -- cycle ;
\draw [line width=1.5]    (410.42,70.59) -- (410.12,109.92) ;
\draw [shift={(410.09,113.92)}, rotate = 270.44] [fill={rgb, 255:red, 0; green, 0; blue, 0 }  ][line width=0.08]  [draw opacity=0] (8.75,-4.2) -- (0,0) -- (8.75,4.2) -- (5.81,0) -- cycle    ;
\draw [line width=1.5]    (196.22,172.04) -- (172.07,149.28) ;
\draw [shift={(169.16,146.54)}, rotate = 43.29] [fill={rgb, 255:red, 0; green, 0; blue, 0 }  ][line width=0.08]  [draw opacity=0] (8.75,-4.2) -- (0,0) -- (8.75,4.2) -- (5.81,0) -- cycle    ;
\draw [line width=1.5]    (298.42,60.42) -- (274.27,37.66) ;
\draw [shift={(271.35,34.92)}, rotate = 43.29] [fill={rgb, 255:red, 0; green, 0; blue, 0 }  ][line width=0.08]  [draw opacity=0] (8.75,-4.2) -- (0,0) -- (8.75,4.2) -- (5.81,0) -- cycle    ;
\draw   (192.75,92.55) .. controls (192.54,89.99) and (194.46,87.75) .. (197.02,87.55) .. controls (199.58,87.35) and (201.82,89.26) .. (202.02,91.83) .. controls (202.22,94.39) and (200.31,96.63) .. (197.75,96.83) .. controls (195.18,97.03) and (192.95,95.11) .. (192.75,92.55) -- cycle ;
\draw    (151.07,49) -- (193.41,88.83) ;
\draw [shift={(149.61,47.63)}, rotate = 43.25] [fill={rgb, 255:red, 0; green, 0; blue, 0 }  ][line width=0.08]  [draw opacity=0] (12,-3) -- (0,0) -- (12,3) -- cycle    ;
\draw    (239.68,45.24) -- (200.04,87.77) ;
\draw [shift={(241.05,43.78)}, rotate = 132.99] [fill={rgb, 255:red, 0; green, 0; blue, 0 }  ][line width=0.08]  [draw opacity=0] (12,-3) -- (0,0) -- (12,3) -- cycle    ;
\draw  [fill={rgb, 255:red, 0; green, 0; blue, 0 }  ,fill opacity=1 ] (196.51,93.1) .. controls (196,92.62) and (195.99,91.82) .. (196.47,91.31) .. controls (196.95,90.81) and (197.75,90.79) .. (198.26,91.27) .. controls (198.77,91.76) and (198.78,92.56) .. (198.3,93.07) .. controls (197.82,93.57) and (197.01,93.59) .. (196.51,93.1) -- cycle ;
\draw  [draw opacity=0] (295.51,113.84) .. controls (295.56,113.84) and (295.62,113.84) .. (295.67,113.84) .. controls (301.93,113.99) and (306.85,120.97) .. (306.66,129.45) .. controls (306.57,133.33) and (305.43,136.85) .. (303.61,139.52) -- (295.32,129.19) -- cycle ; \draw    (295.67,113.84) .. controls (301.93,113.99) and (306.85,120.97) .. (306.66,129.45) .. controls (306.59,132.56) and (305.85,135.43) .. (304.62,137.82) ; \draw [shift={(303.61,139.52)}, rotate = 291.91] [fill={rgb, 255:red, 0; green, 0; blue, 0 }  ][line width=0.08]  [draw opacity=0] (7.2,-1.8) -- (0,0) -- (7.2,1.8) -- cycle    ; \draw [shift={(295.51,113.84)}, rotate = 24.68] [fill={rgb, 255:red, 0; green, 0; blue, 0 }  ][line width=0.08]  [draw opacity=0] (7.2,-1.8) -- (0,0) -- (7.2,1.8) -- cycle    ;

\draw (413.67,79.33) node [anchor=north west][inner sep=0.75pt]   [align=left] {{\small $\vect g$}};
\draw (453,190) node [anchor=north west][inner sep=0.75pt]   [align=left] {{\small $\hat \imath _{\rm{A}}$}};
\draw (375,120) node [anchor=north west][inner sep=0.75pt]   [align=left] {{\small $\hat \jmath _{\rm{A}}$}};
\draw (234,31) node [anchor=north west][inner sep=0.75pt]  [rotate=-313.75] [align=left] {{\small $\hat \imath _{\rm B}$}};
\draw (129.9,40) node [anchor=north west][inner sep=0.75pt]  [rotate=-313.75] [align=left] {{\small $\hat \jmath _{\rm B}$}};
\draw (287,20.33) node [anchor=north west][inner sep=0.75pt]   [align=left] {{\small ${\vect f}_2$}};
\draw (184,135.33) node [anchor=north west][inner sep=0.75pt]   [align=left] {{\small ${\vect f}_1$}};
\draw (260.67,125) node [anchor=north west][inner sep=0.75pt]   [align=left] {{\small c}};
\draw (327.33,200) node [anchor=north west][inner sep=0.75pt]   [align=left] {{\small \textit{w}}};
\draw (308.33,115) node [anchor=north west][inner sep=0.75pt]   [align=left] {{\small $\theta$}};
\end{tikzpicture}
}
\caption{Bicopter configuration considered in this paper. The bicopter is constrained to the {$\hat  \imath _\rmA- \hat \jmath _\rmA$} plane and rotates about the {$\hat k_\rmA$} axis of the inertial frame {$\rm F_A.$}
Note that $\vect r_{\rmc/w} = r_1 \hat \imath_\rmA + r_2 \hat \jmath_\rmA.$}
    \label{fig:Bicopter}
\end{figure}

Next, we consider the bicopter system shown in Figure \ref{fig:Bicopter}.
The derivation of the equations of motion is described in more detail in \cite{bicopter}.
%
The equations of motion are
\begin{align}
    m \ddot r_1 &= -F \sin \theta, 
    \label{eq:eom_r1_2}
    \\
    m \ddot r_2 &= F \cos \theta - m g,
    \label{eq:eom_r2_2}
    \\
    J \ddot \theta &= M,
    \label{eq:eom_theta_2}
\end{align}
where
$r_1$ and $r_2$ are the horizontal and vertical positions of the center of mass of the bicopter, 
$\theta$ is the roll angle of the bicopter, 
$F$ and $M$ are the total force and the total moment applied to the bicopter, respectively,  
and 
$m$ and $J$ are the mass and the moment of inertia of the bicopter, respectively.

\begin{figure}[h!]
    \centering
    \resizebox{\columnwidth}{!}
    {
    \begin{tikzpicture}[auto, node distance=2cm,>=latex',text centered]
    
        \node [smallblock, minimum height=3em, text width=1.6cm] (Mission) { Mission Planner};
        \node [smallblock, minimum height=3em, right = 7em of Mission, text width=1.6cm] (Pos_Cont) { Position Controller};

        \node [right = 3 em of Pos_Cont] (midpoint) {};
        
        \node [smallblock, minimum height=3em, below = 1.5em of midpoint,text width=1.6cm] (Att_Cont) {Attitude Controller};
        \node [smallblock, minimum height=3em, minimum width = 5.5em,  right = 7em of midpoint] (Quadcopter) { bicopter};
        
        \draw [blue,->] (Mission) -- node[above, xshift = -0.05 em]{Position setpoint} (Pos_Cont);
        \draw[red,->] (Mission.-90) |- +(0,-.5) 
        node[xshift = 6 em, yshift = 0.25 em]{Roll setpoint}
        |- (Att_Cont.180);
        %
        

        \draw [red,->] (Pos_Cont.15) -| (Att_Cont.90);
        
        \draw [blue,->] (Pos_Cont.15) -- node  [xshift = 0em, above] {Force} (Quadcopter.165);
        
        \draw [red,->] (Att_Cont.0) -- +(0.1,0) |-  node [below, xshift = 2 em]{Moment}(Quadcopter.195);
        \draw [blue,->] (Quadcopter.15) -- +(1,0) |- node[below,xshift = -10em, above]{Position and Velocity Measurement} ([yshift = -7.5 em]Pos_Cont.south) -- (Pos_Cont.south);
        \draw [red,->] (Quadcopter.345) -- +(0.5,0) |- node[above,xshift = -7em]{Attitude and Rate Measurement} ([yshift = -2 em]Att_Cont.south) -- (Att_Cont.south);

    \end{tikzpicture}
    }
    \caption{Bicopter Control Architecture. }
    \label{fig:autopilot_nested_loop}
\end{figure}
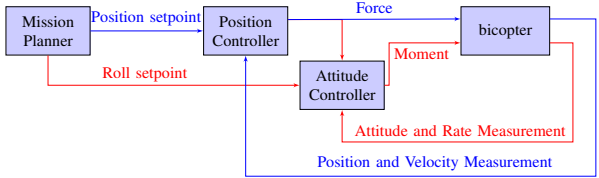

In this work, we consider a cascaded architecture for the bicopter tracking problem.
The cascaded architecture is shown in Figure \ref{fig:autopilot_nested_loop}.
The cascaded loop architecture is motivated by the time separation principle, which requires the inner loop dynamics to be significantly faster than the outer loop dynamics. 
The outer loop is designed to track the position references and the inner loop is designed to track the angle references.
%
%
With the faster rotational dynamics, the two-dimensional force vector required to move the bicopter in a desired direction can be realized by first adjusting the angle of the bicopter so that the axis of the bicopter is along the direction of the force vector and then modulating the total force exerted by the propellers.

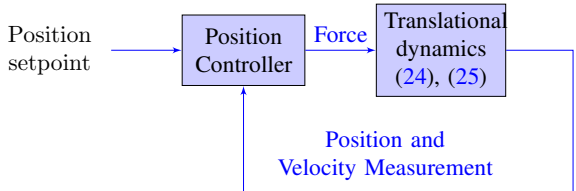
\begin{figure}[h!]
    \centering
    \resizebox{\columnwidth}{!}
    {
    \begin{tikzpicture}[auto, node distance=2cm,>=latex',text centered]
    
        \node (Mission) 
        {
            $\begin{array}{c}
                \rm Position \\
                \rm setpoint 
            \end{array}$            
        };
        \node [smallblock, minimum height=3em, right = 3em of Mission, text width=1.6cm] (Pos_Cont) { Position Controller};

        \node [smallblock, minimum height=3em, minimum width = 5.5em,  right = 3em of Pos_Cont,  text width=1.7cm] (Quadcopter) { Translational dynamics \eqref{eq:r1_outer_loop_eqn}, \eqref{eq:r2_outer_loop_eqn}};
        
        \draw [blue, ->] (Mission) -- (Pos_Cont);

        \draw [blue,->] (Pos_Cont) -- node  [xshift = 0em, above] {Force}  (Quadcopter.180);

        \draw [blue,->] (Quadcopter.0) -- +(1,0) |- node[below,xshift = -8em, above]
        {
            $\begin{array}{c}
                \text{Position and} \\
                \text{Velocity Measurement}
            \end{array}$            
        }
        ([yshift = -4.5 em]Pos_Cont.south) -- (Pos_Cont.south);

    \end{tikzpicture}
    }
    \caption{Outer loop to track position setpoints. }
    \label{fig:outer_loop}
\end{figure}

Since the inner loop, which regulates the rotational motion, is assumed to be significantly faster than the outer loop, the translational motion of the bicopter satisfies
\begin{align}
    m \ddot r_1 &= f_{r_1},
    \label{eq:r1_outer_loop_eqn}
    \\
    m \ddot r_2 &= f_{r_2},
    \label{eq:r2_outer_loop_eqn}
\end{align}
where $f_{r_1}$ and $f_{r_2}$ are the horizontal and vertical forces on the bicopter. 
Note that \eqref{eq:r1_outer_loop_eqn}, \eqref{eq:r2_outer_loop_eqn} are linear and decoupled, where the \textit{effective} forces 
%
%
$f_{r_1} $ and $ f_{r_2}$ are assumed to be realizable due to the faster rotational dynamics.
The objective of the outer loop is thus to compute the force input to track a desired position reference, as shown in Figure \ref{fig:outer_loop}.
The position controller considered in this work consists of two decoupled controllers, each tasked to follow the corresponding position command.

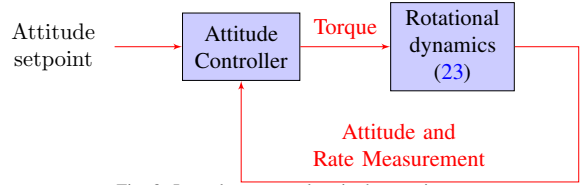
\begin{figure}[h!]
    \centering
    \resizebox{\columnwidth}{!}
    {
    \begin{tikzpicture}[auto, node distance=2cm,>=latex',text centered]
    
        \node (Mission) 
        {
            $\begin{array}{c}
                \rm Attitude \\
                \rm setpoint 
            \end{array}$            
        };
        \node [smallblock, minimum height=3em, right = 3em of Mission, text width=1.6cm] (Pos_Cont) { Attitude Controller};

        \node [smallblock, minimum height=3em, minimum width = 5.5em,  right = 4em of Pos_Cont,  text width=1.6cm] (Quadcopter) { Rotational dynamics \eqref{eq:eom_theta_2}};
        
        \draw [red,->] (Mission) -- (Pos_Cont);

        \draw [red,->] (Pos_Cont) -- node  [xshift = 0em, above] {Torque}  (Quadcopter.180);

        \draw [red,->] (Quadcopter.0) -- +(1,0) |- node[below,xshift = -8em, above]
        {
            $\begin{array}{c}
                \text{Attitude and} \\
                \text{Rate Measurement}
            \end{array}$            
        }
        ([yshift = -4.5 em]Pos_Cont.south) -- (Pos_Cont.south);

    \end{tikzpicture}
    }
    \caption{Inner loop to track attitude setpoints. }
    \label{fig:inner_loop}
\end{figure}

The inner loop computes the moment required to realize the force vector desired by the outer loop by orienting the bicopter such that the force axis of the bicopter aligns with the desired force vector given by the outer loop. 
Note that the desired roll angle is thus given by 
\begin{align}
    \theta_\rmd = \text{atan2}(-f_{r_1}, f_{r_2}).
\end{align}
and the rotational motion of the bicopter is governed by \eqref{eq:eom_theta_2}.
Figure \ref{fig:inner_loop} shows the inner loop to track the desired roll angle.

To follow the position and angle references in the outer and the inner loop, we consider the tracking architectuers described in Section \ref{sec:tracking}.
Note that there are three controllers, two for position tracking and one for angle tracking. 
In each case, the controller can be written as \eqref{eq:u_para},
where the data regressor $\Phi(t)$ is constructed using the measured data and $\theta(t)$ is synthesized by CTRCAC.

To simulate the bicopter in this work, we set $m = 1.5$ $\rm kg$ and $J = 0.03$ $\rm kg m^2.$
We command the bicopter to follow an inclined elliptical trajectory given by
\begin{align}
    r_{\rmd1}(t) &= 5 \cos(\phi)-5 \cos(\phi) \cos (\omega t) - 3 \sin(\phi)\sin(\omega t), \nn\\ 
    r_{\rmd2}(t) &= 5\sin(\phi) - 5 \sin(\phi) \cos (\omega t) + 3 \cos(\phi)\sin(\omega t), \nn
\end{align}
where {$\phi=45~\rm{deg}$ and $\omega = 0.1 \ \rm rad/s^{-1}. $}

\begin{figure}[h]
    \centering
    \includegraphics[width=\columnwidth]{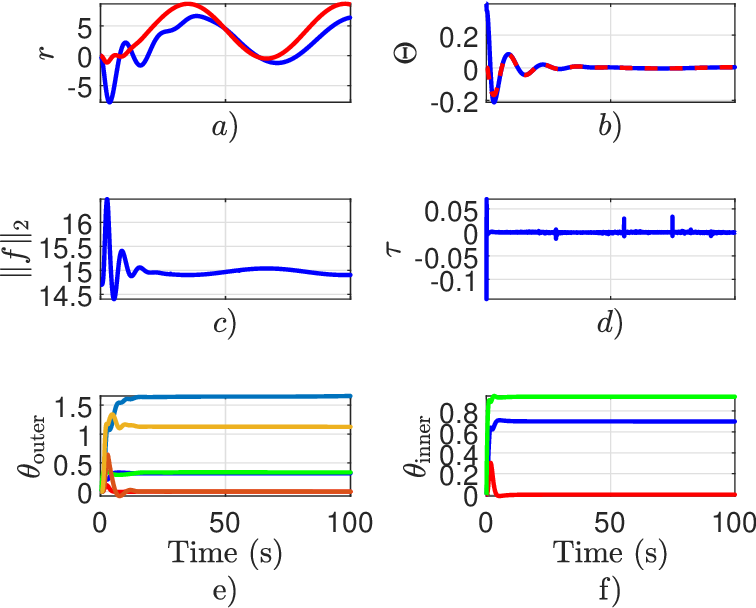}
    \caption{\textbf{Tracking response with PID controller in each loop}. 
    a) position response, 
    b) angle response, 
    c) the magnitude of the force,  
    d) torque applied to the bicopter, 
    e) controller gains in the outer loop, and 
    f) controller gains in the inner loop. 
    }
    \label{fig:bicopter_trajectory_data_pid}
\end{figure}

First, we use a PID controller in each loop to track the desired trajectory. 
Since the translational dynamics \eqref{eq:r1_outer_loop_eqn}, \eqref{eq:r2_outer_loop_eqn} and the rotational dynamics \eqref{eq:EulerEq} are double integrators, we use the same RCAC hyperparameters optimized by PSO for a double integrator in the previous section.
Figure \ref{fig:bicopter_trajectory_data_pid} shows the 
a) position response, 
b) angle response, 
c) the magnitude of the force,   
d) torque applied to the bicopter, 
e) controller gains in the outer loop, and 
f) controller gains in the inner loop. 

\begin{figure}[h]
    \centering
    \includegraphics[width=\columnwidth]{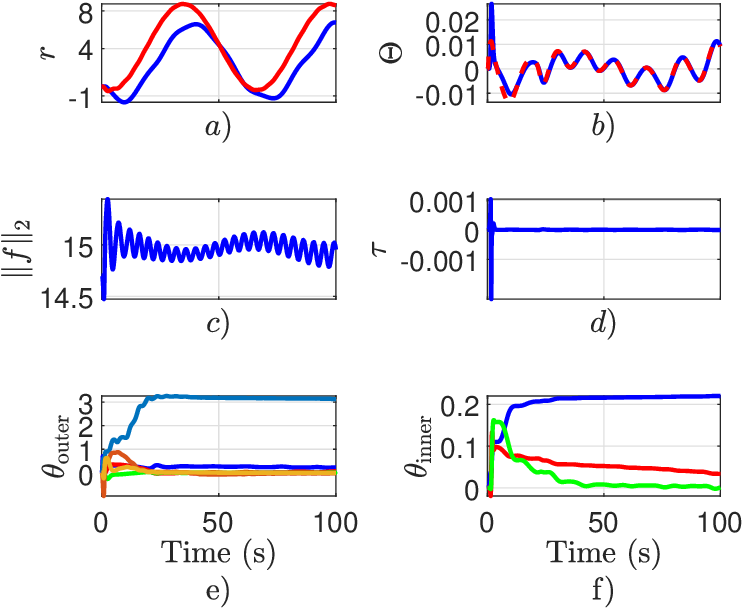}
    \caption{\textbf{Tracking response with modified PPI controller in each loop}. 
    a) position response, 
    b) angle response, 
    c) the magnitude of the force,  
    d) torque applied to the bicopter, 
    e) controller gains in the outer loop, and 
    f) controller gains in the inner loop. 
    }
    \label{fig:bicopter_trajectory_data_ppi}
\end{figure}

Next, we use the cascaded architecture described in Appendix \ref{sec:tracking} in each loop to track the desired trajectory. 
We use the same RCAC hyperparameters optimized by PSO for a double integrator in the previous section.
Figure \ref{fig:bicopter_trajectory_data_ppi} shows the 
a) position response, 
b) angle response, 
c) the magnitude of the force,   
d) torque applied to the bicopter, 
e) controller gains in the outer loop, and 
f) controller gains in the inner loop. 

\begin{figure}[h]
    \centering
    \includegraphics[width=200pt]{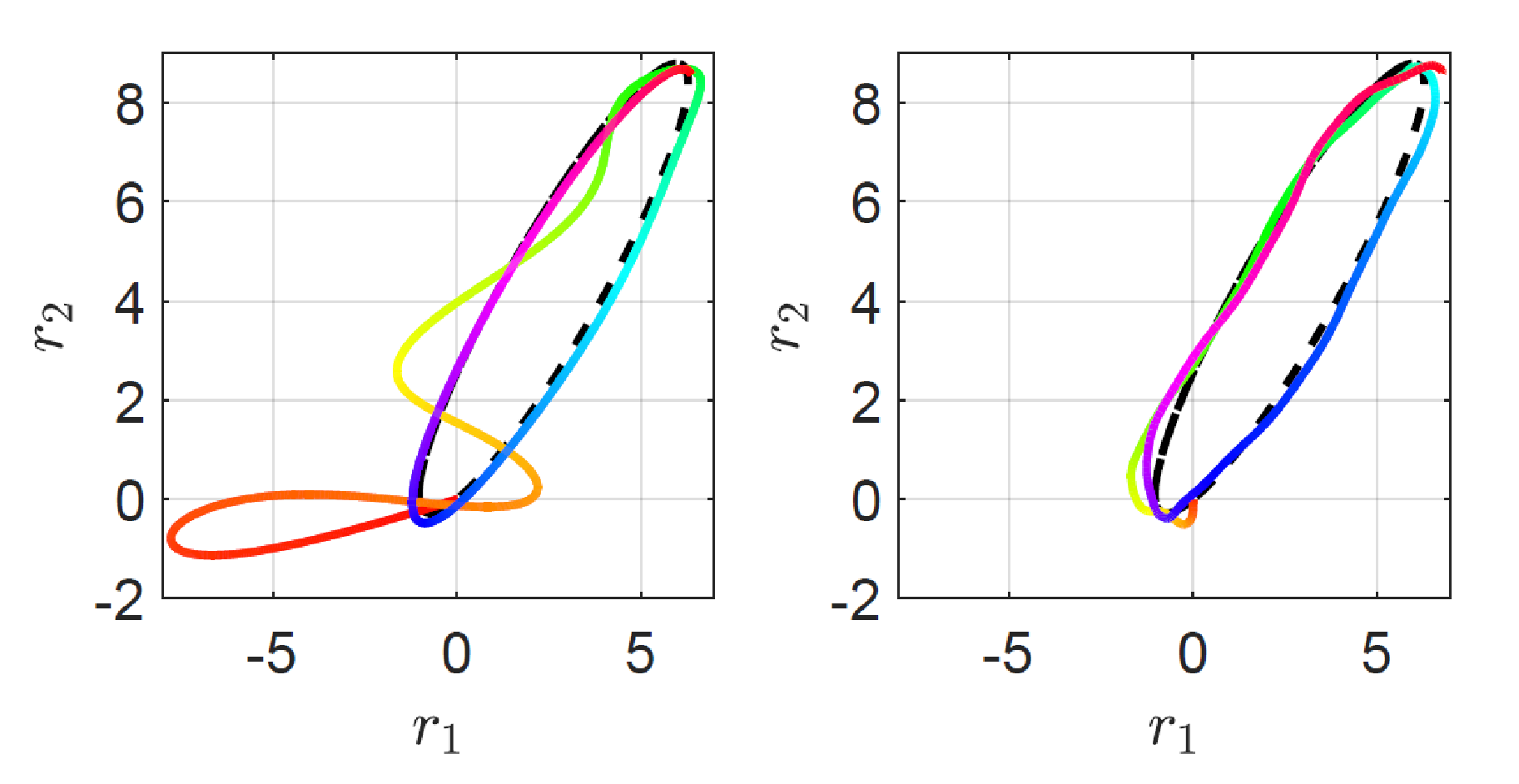}
    \caption{Eliptical trajectory tracking  response with adaptive PID controllers (on the left) and modified PPI controllers (on the right).}
    \label{fig:bicopter_trajectory_compare}
\end{figure}

Figure \ref{fig:bicopter_trajectory_compare} shows the trajectory tracking response with the  PID and the modified adaptive PPI controller.
In several numerical experiments with various RCAC hyperparameters, it is observed that the trajectory tracking response obtained with modified adaptive PPI controllers, in general, has better transient characteristics than the response obtained with adaptive PID controllers.

%
\subsection{Three-dimensional Attitude Control}

Next, we consider the 3-dimensional attitude control problem. 
For a rigid body, the attitude dynamics is 
\begin{align}
    \dot \Theta &= S(\Theta) \omega, 
    \label{eq:PoissonEq}
    \\
    J \dot \omega + \omega \times J w &= \tau, 
    \label{eq:EulerEq}
\end{align}
where $\Theta \in \BBR^3$ is an Euler angle sequence that parameterizes the attitude of the rigid body with respect to an inertial frame, 
$S(\Theta) \in \BBR^{3 \times 3}$ is a matrix that depends on the choice of Euler angle sequence,  
$\omega \in \BBR^3$ is the angular velocity vector of the rigid body relative to an inertial frame measured in a body-fixed frame, 
$J \in \BBR^{3 \times 3}$ is the momemt of the inertia matrix of the rigid body and $\tau$ is the moment applied to the rigid body and measured in the body-fixed frame. 
Note that \eqref{eq:PoissonEq} is the Poisson's equation, written in terms of the Euler angles and \eqref{eq:EulerEq} is the Euler's equation governing the rotational dynamics \cite{kasdin2011engineering}. 
The model-free attitude control problem objective is to synthesize a stabilizing and tracking controller for the Euler angles of a rigid body without using \eqref{eq:PoissonEq}, \eqref{eq:EulerEq}. 

\begin{figure}[h]
    \centering
    \includegraphics[width=\columnwidth]{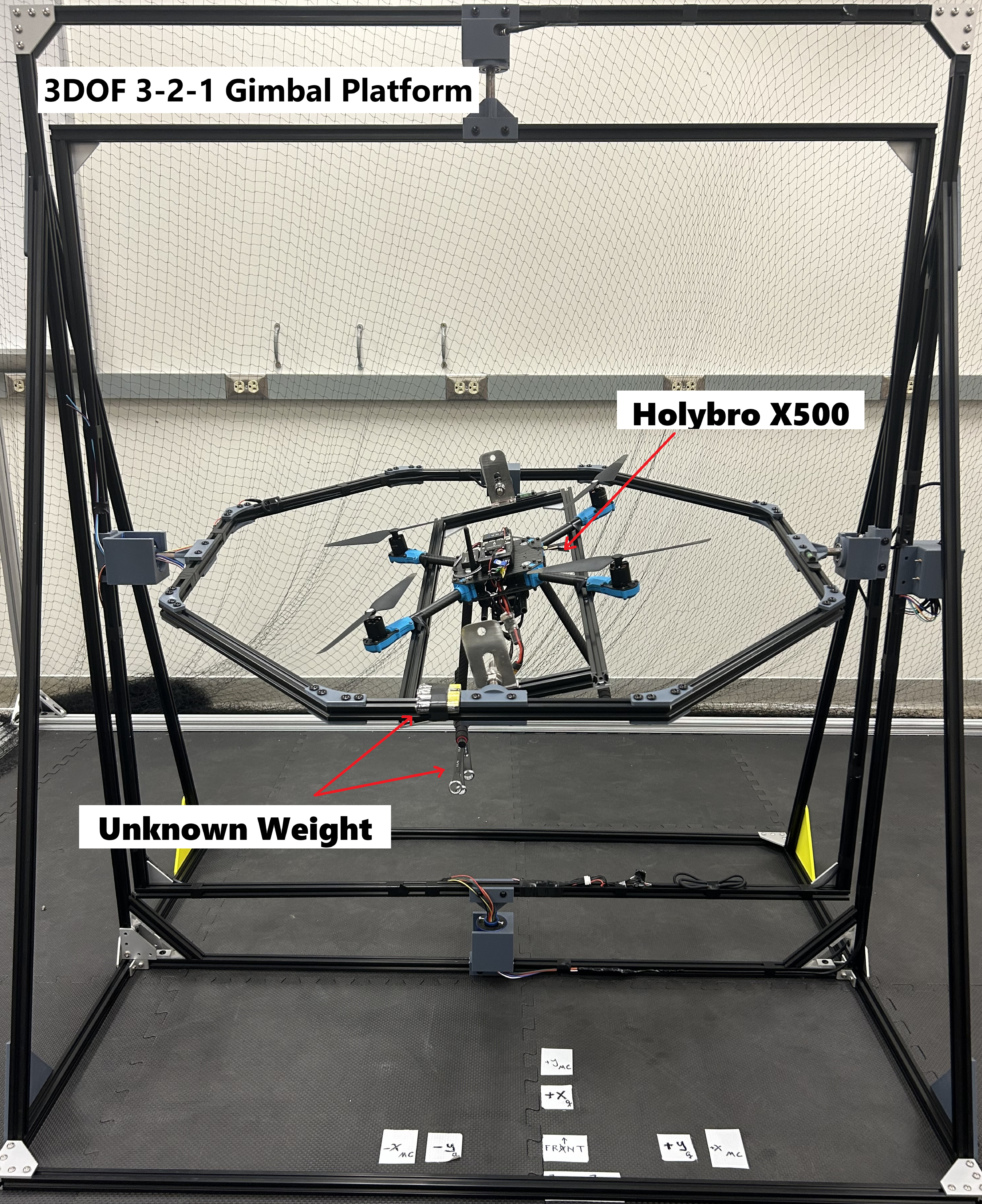}
        \caption{Experimental set up for model-free attitude control of Holybro X500 quadcopter frame. The three gimbal rings allow independent 3-2-1 rotation of the quadcopter frame. Unknown weights are added to the middle ring and the quadcopter frame to simulate unknown external disturbances.}
    \label{fig:3d_platform_pic}
\end{figure}

In this work, we consider the attitude stabilization problem of the Holybro X500 quadcopter airframe mounted on a 3dof 3-2-1 gimbal platform shown in Figure \ref{fig:3d_platform_pic}. 
The three gimbal rings allow independent 3-2-1 rotation of the quadcopter frame. 
Two unknown weights are added to the frame to reflect unknown constant disturbance torques in the roll and pitch axis. 

\begin{figure}[h]
    \centering
    \includegraphics[width=\columnwidth]{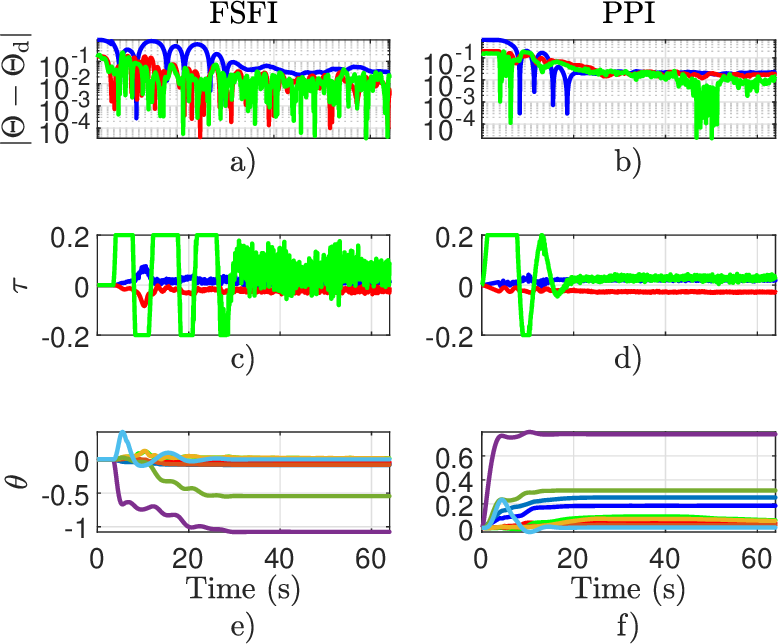}
        \caption{Holybro X500 airframe attitude stabilization with the FSFI and the cascaded  control architectures.
        a), b) show the Euler angle errors, 
        c), d) show the torque signals given by the controller, and 
        e), f) show the controller gains updated CTRCAC for the FSFI architecture and the cascaded PPI architecture. 
       }
    \label{fig:platform_exp}
\end{figure}

Three independent controller loops are implemented to stabilize each of the 3-2-1 Euler angles, where the controllers in each loop are updated by CTRCAC.
To stabilize the attitude of the X500 airframe, we use the FSFI\footnote{\href{https://www.youtube.com/watch?v=SNRrs-B\_v0M}{https://www.youtube.com/watch?v=SNRrs-B\_v0M}} 
and the cascaded controller\footnote{\href{https://www.youtube.com/watch?v=XgQvo6G7K_E}{https://www.youtube.com/watch?v=XgQvo6G7K\_E}} architecture to stabilize each Euler angle.
In RCAC, we set $P_0 = R_z = 10^4$, and $ p_f = 2. $
The magnitude of the torque signals computed by the adaptive controllers is saturated at 0.2 for safety. 
Figure \ref{fig:platform_exp} shows the Euler angle errors, the torque signals given by the controller, and the controller gains updated CTRCAC for the FSFI architecture and the cascaded PPI architecture.



\section{Conclusion}
\label{sec:conclusions}

This paper extended the continuous-time retrospective cost adaptive control to multi-input, multi-output systems, introduced state-space filtering, and verified its application in both a simulation and an experimental framework. 
%
%
In particular, the adaptive controller is applied to stabilize a double integrator and a bicopter system in a simulation framework and the Holybro X500 quadcopter airframe in an experimental framework. 

Four control architectures are considered for the stabilization and tracking control problems.
In particular, an $n$-th order transfer function in a basic servo loop, 
a PID controller in the basic servo loop, 
a cascaded architecture, and 
a full state feedback control architecture with integral action are considered. 
CTRCAC is used to update the controllers in each of these architectures. 
In the simulation framework, a particle swarm optimizer is used to tune the CTRCAC hyperparameters, whereas the hyperparameters are manually tuned in the experiment.  
In all of the scenarios considered, CTRCAC yields stabilizing and tracking controllers with minimal hyperparameter tuning.


\bibliographystyle{IEEEtran}
\bibliography{IEEEabrv,bib/b,bib/PX4bib}

\renewcommand{\thesection}{\Alph{section}}
\section*{Appendix}
\setcounter{section}{0}
\section{Controller Parameterization}
\label{appndx:ControllerPara}
This appendix briefly describes the construction of the regressor for the linear parameterization of an $n$-th order transfer function and a PID control law. 


Consider the $n$-th order SISO transfer function
\begin{align}
    G(s)
        =   
            \frac
                {b_{n-1}s^{n-1} + \ldots + b_0}
                {s^{n} + a_{n-1}s^{n-1} + \ldots + a_0},
    \label{eq:u_TF}
\end{align}
where, for $j = 1,2, \ldots, n, $ $a_j, b_j \in \BBR.$
Let $z$ denote the output error and $u$ denote the input of the transfer function $G(s).$
Then, 
\begin{align}
    u(t) = \phi(t) \theta, 
\end{align}
where 
\begin{align}
    \phi(t)
        \isdef
            \matl 
                -\int u & \ldots -\int^{(n)} u \quad 
                \int z & \ldots \int^{(n)} z
            \matr, 
\end{align}
and 
$
    \theta
        \isdef
            \matl 
                a_{n-1}&
                \cdots &
                a_0&
                b_{n-1} &
                \cdots &
                b_0
            \matr^\rmT.
$
Note that
\begin{align}
    \int^{(j)} u
        \isdef             
            \underset{j \ \text{times}}{\operatorname{\int \ldots \int }} 
            u(\tau_j) 
            \rmd \tau_j 
            \rmd \tau_{j-1}
            \ldots 
            \rmd \tau_1 .
\end{align}


    
    


Consider the PID controller
\begin{align}
    u(t) 
        = 
            k_\rmp z(t) +
            k_\rmi \int_0^t z(\tau) \rmd \tau +
            k_\rmd \dot z(t),
    \label{eq:u_PID}
\end{align}
where $k_\rmp, k_\rmi,$ and $k_\rmd$ are the proportional, integral, and the derivative gains, respectively. 
Note that the control law \eqref{eq:u_PID} can be written as
\begin{align}
    u(t) = \phi(t) \theta, 
\end{align}
where 
\begin{align}
    \phi(t)
        \isdef
            \matl 
                z(t) & 
                \int_0^t z(\tau) \rmd \tau
                & \dot z(t)
            \matr, 
\end{align}
and 
$
    \theta
        \isdef
            \matl 
                k_\rmp &
                k_\rmi &
                k_\rmd
            \matr^\rmT.
$

\section{Tracking Control Architecture}
\label{sec:tracking}
This appendix briefly reviews the control system architectures for tracking problems. 

%

\begin{figure}[H]
    \centering
    \resizebox{\columnwidth}{!}{
    \begin{tikzpicture}[auto, node distance=1cm ]
    
        \node [input, name=input] {};
        \node [sum, right of=input, name=sum1, node distance = 1cm] {+};
        \node [block,minimum width=1cm, right of=sum1, name=PCONTROL, node distance = 1.25cm] {P};
        \node [output, right of=PCONTROL,node distance = 0.75cm] (outputp) {};
        \node [sum, right of=PCONTROL,node distance = 1.25cm ,name=sum2] {+};
        \node [block,minimum width=1cm, right of=sum2, name=PICONTROL, node distance = 1.25cm] {PI};
        
        \node [sum, right of=PICONTROL, name=sum3, node distance = 1.25cm] {+};
        \node [block,minimum width=1cm, right of=sum3, name=PLANT, node distance = 1.25cm] {Plant};
        \node [output, right of=PLANT,node distance = 1cm] (output1) {};
        \node [input, below of=sum1, name=undersum1, node distance = 1.25cm] {};
        \node [input, below of=sum2, name=undersum2] {};
        \node [input, above of=sum3] (abovesum3) {};
    
        \draw [->] (input) -- node {$r$} (sum1);
        \draw [->] (sum1) -- node {$z$} (PCONTROL);
        \draw [->] (PCONTROL) -- (sum2);
        \draw [->] (sum2) -- node {} (PICONTROL);
        \draw [->] (PICONTROL) -- (sum3);
        \draw [->] (sum3) -- node {$u$} (PLANT);
        \draw [-] (PLANT) -- (output1);
        \draw [-] (PLANT.0) -| +(0.25,-0.5) |- (undersum1);
        \draw [-] (PLANT.320) -| +(0.125,-0.5) |- (undersum2);
        \draw [->] (undersum1) -- node [xshift = 1.75em] {$-y$}(sum1);
        \draw [->] (undersum2) -- node [xshift = 1.75em] {$-\dot y$} (sum2);
        \draw [-] (outputp) |- (abovesum3);
        \draw [->] (abovesum3) -| (sum3);

    \end{tikzpicture}
    }
    \caption{Modified cascaded PPI with control architecture for command following.}
    \label{fig:cascaded-ppi}
    \vspace{-1em}
\end{figure}
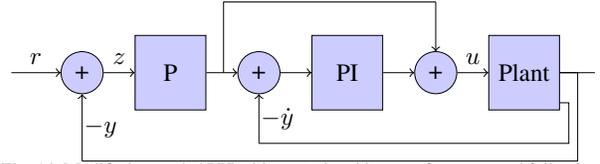

Figure \ref{fig:cascaded-ppi} shows the modified cascaded PPI architecture.
The PPI architecture is typically used for plants in strict feedback form and allows the imposition of desired transient characteristics to the outer loop variable. 

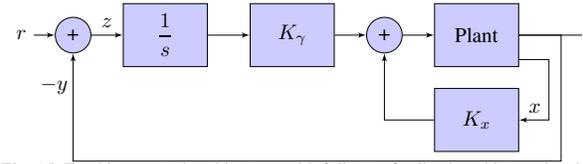
\begin{figure}[h!]
    \centering
    \resizebox{\columnwidth}{!}{
    \begin{tikzpicture}[auto, node distance=2cm,>=latex',text centered]
    
        \node at (-3,0) (reference) {$r$};
        \node[sum, right = 1. em of reference] (sum0) {+};
        \node[draw = none] at (sum0.center) {};
        \node [smallblock, right = 1.5 em of sum0] (integrator) {$\dfrac{1}{s}$};
        
        \node [smallblock, right = 2 em of integrator] (IntGain) {$K_\gamma$};
        \node[sum, right = 1.5 em of IntGain] (sum1) {+};
        \node[draw = none] at (sum1.center) {};
        
        \node [smallblock, right = 1.5 em of sum1] (Plant) {Plant};
        \node [smallblock, below = 1 em of Plant] (FSFG) {$K_x$};
        \node[right = 3 em of Plant] (output) {};
        
        \draw[->] (reference) -- (sum0);
        \draw[->] (sum0) -- node[above]{$z$}(integrator);
        \draw[->] (integrator) -- node[above]{} (IntGain);
        \draw[->] (IntGain) -- (sum1);
        \draw[->] (sum1) -- (Plant);
        \draw[-] (Plant) --  (output);
        \draw[->] (Plant.330) -- +(0.5,0) |- (FSFG.0) node[xshift = 0.75em, above]{$x$};
        \draw[->] (FSFG.west) -| (sum1.south);
        \draw[->] (Plant.east) -| ([xshift = 2em, yshift = -6em]Plant.east) -| node[near end, xshift = 0.1em, yshift = 1em]{$-y$} node[near end, xshift = 1.5em, yshift = -1.25em]{}(sum0.south);
        
    \end{tikzpicture}
    }
        \caption{Tracking control architecture with full state feedback and integral action.}
        \label{fig:FSF_control_loop_int}
\end{figure}

Figure \ref{fig:FSF_control_loop_int} shows the full state feedback control architecture with integral action (FSFI) consists of a full state feedback controller in the inner loop for stabilization and an integrator in the outer loop for command following.

The stability analysis of these two architectures is presented in \cite{chee2024performance}.

\end{document}